\documentstyle[aps,prl,multicol,epsf]{revtex}
\input epsf
\begin{document}
\draft

\begin{multicols}{2}

\noindent
{\bf Comment on: ``Pipe Network Model for Scaling of
Dynamic Interfaces in Porous Media''}

In a recent Letter \cite{Lam_00}, Lam and Horv\a'ath (LH) present a
pipe network model of imbibition.  The network was made of small
pipes uniformly distributed in sizes and large pipes of a single radius.
By tuning the size and relative number of large to small pipes
LH were able to find power law exponents
close to those measured in the experiment of one of the
authors\cite{Horvath_95}.  For example
model parameters could be chosen
such that the mean velocity of the front ($v$) is related
to the mean height of the front ($H$) as $v\sim H^{-\Omega}$ with
$\Omega = 1.6$.  A new exponent identity is proposed
(i.e., Eq. (7) of the Letter) relating temporal and spatial correlation
functions.  In this comment we point out that LH neglect an important
length scale and that their exponent identity, whose
underlying assumption holds trivially for
standard kinetic roughening, should not apply to imbibition.

To explain the main flaw in LH's arguments we recall the
picture of spontaneous imbibition developed in Refs.
\cite{Dube_99,Dube_00}.  Competition between the effective
line tension and liquid transport imposes to the front
a lateral length scale $\xi_{\times}$ {\it regardless} of
the value of $\Omega$ (indeed, in experiments using paper as the
porous material a variety of $\Omega$'s different from unity and
a variety of scaling behaviors have been reported \cite{Dube_00}).
$\xi_{\times}$ is important since for any fixed
height the width of the interface is finite and does not obey
a power law relationship with system size or time.
The width only diverges if the
average height of the interface diverges as is the case for
freely rising fronts.
For example in the model of Ref. \cite{Dube_99} the width
follows $w \sim \xi_{\times}^\chi$ with a {\em global} roughness
exponent $\chi = 1.25$ and the lateral correlation
length scales as $\xi_{\times} \sim v^{-1/2} \sim t^{1/4}$.

The existence of an intrinsic lateral correlation length becomes
obvious when systems of different sizes are considered.
Although such an analysis was not presented by LH a lateral
correlation length can be inferred from their description
of the height difference correlation function given as;
\begin{equation}
  C(l) = v^{-\kappa} g(l v^{(\theta_l+\kappa)/\alpha}),
\label{cl}
\end{equation}
where $g(x) \sim x^\alpha$ for $x \ll 1$, and constant
for large $x$. The reported values of the
exponents are $\kappa = 0.49$, $\alpha = 0.61$, and $\theta_l =
-0.25$.  Equation (\ref{cl}) clearly
defines a length scale $\xi_v \sim
v^{-\gamma}$, where $\gamma \equiv (\theta_l + \kappa)/\alpha
\simeq 0.4$, which then implies $C(l) \sim \xi_v^{\chi}
g(r/\xi_v)$. Thus, $\xi_v \sim v^{-0.4}$ is analogous to the $\xi_{\times}
\sim v^{-1/2}$ in Ref. \cite{Dube_99}.  Equation (\ref{cl}) also
defines a {\em global} roughness exponent,
$\chi = \kappa \alpha /(\kappa + \theta_l)=1.25$, which
implies a ``superrough'' interface with anomalous
scaling\cite{Lopez_97}, as could be verified from the
structure factor $S (k,t) \equiv \langle | h_k (t) |^2 \rangle$.

To derive their exponent identity, LH assume that the dynamics
are controlled by a single time scale which can be obtained by
two means. For a moving interface, a width $w$ is
reached after a time $t_1 \sim w^{(\Omega+1) / (\kappa \Omega)}$.
On the other hand, another time scale can be obtained from the
interfacial
fluctuations when the average interface height above a reservoir
is kept fixed. In the steady state, LH write
the time correlation function in the form
\begin{equation}
\label{ct}
C(t) = v^{-\kappa} f(t v^{(\theta_t + \kappa)/\beta})
\end{equation}
with the scaling function $f(x) \sim x^{\beta}$ for $x \ll 1$
 and constant at large $x$. This suggests a time scale $t_2
\sim v^{-(\theta_t + \kappa)/\beta} \sim
w^{(\theta_t+\kappa)/\kappa \beta}$, where $\theta_t = 0.38$ and
$\beta = 0.63$ are found.  Assuming that $t_1 \propto t_2$,
LH obtain $\beta = \Omega(\theta_t+
\kappa)/(\Omega+1) \approx 0.54$ which differs from the
direct numerical fit quoted above by $\sim$ 14\%.

Our main objection to the new exponent identity
is that it lacks justification since $t_1$ and
$t_2$ describe two distinct physical time
scales. It takes a time $t_1$ for the
correlation length to have value $\xi_v (t_1)$. This in turns
controls the width as $w \sim \xi_v^{\chi}$. On the other hand,
$t_2$ is the relaxation time of
the fluctuations when the interface is kept at a
fixed height. In this case, the correlation length $\xi_v$ is
a predetermined constant and the saturation within this zone
is obtained when the spatial extent of the correlations equals
$\xi_v$.
The scenario proposed by LH cannot occur if $t_2 \ll t_1$ and
indeed their data indicate $t_2 \sim w^{2.8}$ and $t_1 \sim w^{3.25}$
which shows their assumption to be false asymptotically.
Notice that
the exponents are close, so 
a reasonable range of scaling is needed
to see the difference clearly.

\noindent
M.\ Dub\a'e, \\
{\small Centre for the Physics of Materials, McGill University,
Montr\a'eal, Canada H3A 2T8.}

\noindent
M.\ Rost, \\
{\small Institut f\"ur Theoretische Physik,
Universit\"at zu K\"oln,
D-50937 K\"oln, Germany.}

\noindent
K.R.\ Elder, \\
{\small Dept. of Physics, Oakland University, Rochester, MI
48309}

\noindent
M.\ Alava$^1$, S.\ Majaniemi$^{1,2}$
and T.\ Ala-Nissila$^{1,2}$, \\
{\small $^{1}$Laboratory of Physics and
$^{2}$Helsinki Institute of Physics,
Helsinki University of Technology, FIN-02015 HUT,
Finland.}



\end{multicols}

\end{document}